\documentclass[aps,prl,twocolumn,superscriptaddress,altaffillsymbol]{revtex4-2}

\usepackage[autostyle=true]{csquotes}

\usepackage{amsmath}
\usepackage{amsfonts}
\usepackage{mathtools}
\usepackage{amsbsy}
\usepackage{graphicx}
\usepackage{mathtools}

\usepackage{hyperref}
\hypersetup{unicode=true,
bookmarks=true,bookmarksnumbered=true,bookmarksopen=true,bookmarksopenlevel=2,
breaklinks=false,pdfborder={0 0 1},backref=false,colorlinks=true,allcolors=blue}

\begin{document}

\title{Planar chirality and optical spin-orbit coupling for chiral Fabry-Perot cavities}

\author{J\'er\^ome Gautier}
\author{Minghao Li}
\author{Thomas W. Ebbesen}
\author{Cyriaque Genet}
\email{genet@unistra.fr}
\affiliation{Universit\'e de Strasbourg, CNRS, Institut de Science et d'Ing\'enierie Supramol\'eculaires, UMR 7006, F-67000 Strasbourg, France}

\begin{abstract}
We design, in a most simple way, Fabry-Perot cavities with longitudinal chiral modes by sandwiching between two smooth metallic silver mirrors a layer of polystyrene made planar chiral by torsional shear stress. We demonstrate that the helicity-preserving features of our cavities stem from a spin-orbit coupling mechanism seeded inside the cavities by the specific chiroptical features of planar chirality. Planar chirality gives rise to an extrinsic source of three-dimensional chirality under oblique illumination that endows the cavities with enantiomorphic signatures measured experimentally and simulated with excellent agreement. The simplicity of our scheme is particularly promising in the context of chiral cavity QED and polaritonic asymmetric chemistry. 
\end{abstract}

\maketitle

The design of chiral cavities with modes preserving optical helicity has recently become a major goal in the field of light-matter interactions. Coupling matter to chiral optical modes indeed enriches the field with symmetry-breaking effects that have a great potential,  
\begin{figure}[htb]
  \centering{
    \includegraphics[width=0.9\linewidth]{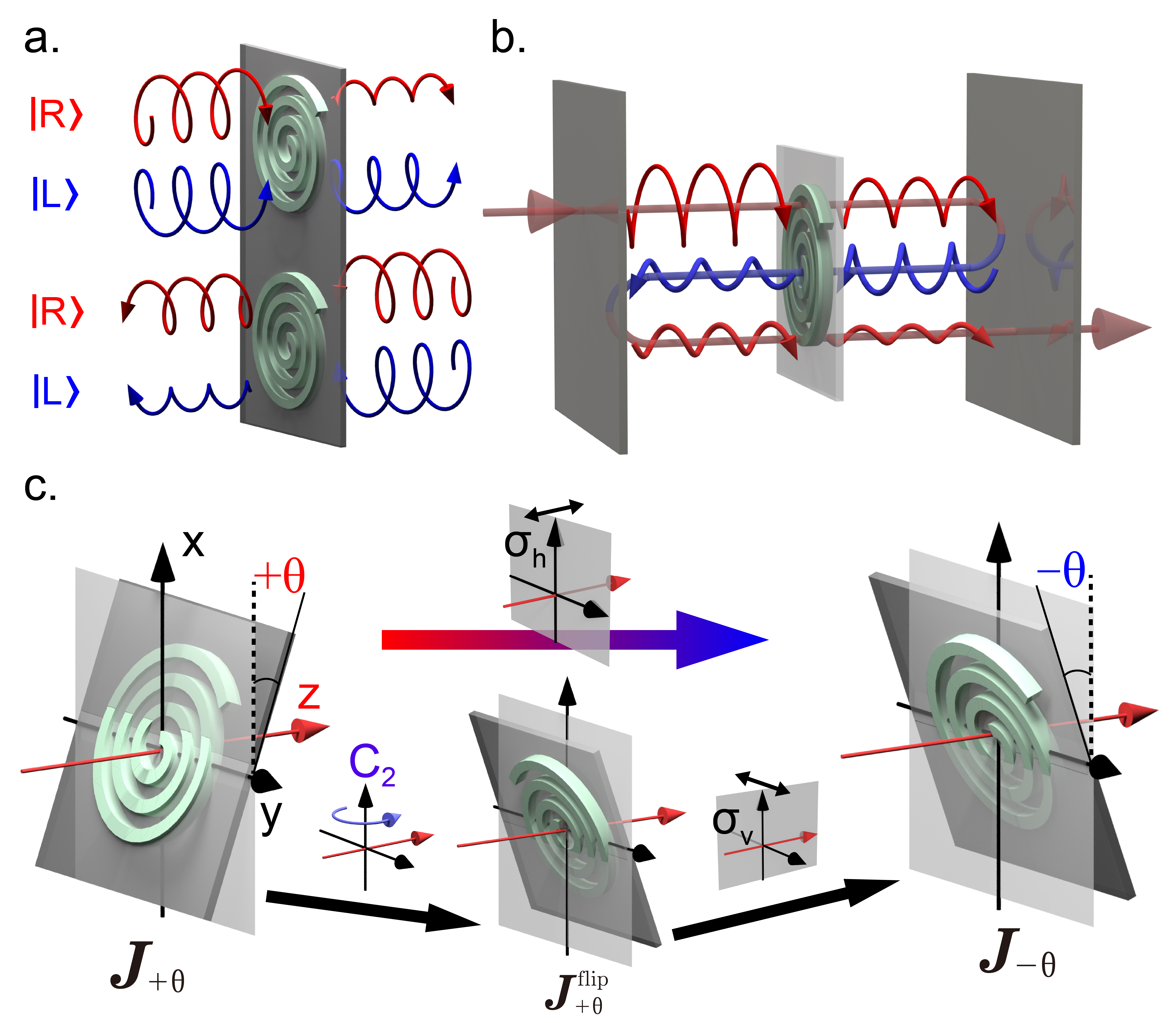}}
    \caption{Panel (a) schematizes the spin-orbit coupling mechanism at play through a planar chiral polymer system -here conceptually represented by a 2D spiral. Planar (2D) chirality (in its most general form, that is without any rotational invariance --see \cite{Drezet2008}) is characterized by circular polarization conversions that depend on the direction of the probe beam transmitted through the spiral. Panel (b) illustrates the breaking of left- vs. right-handed polarization in a Fabry-Perot cavity composed of two usual metallic mirrors but enclosing a 2D chiral medium.  Panel (c) describes how a planar chiral system viewed under oblique illumination yields signatures of 3D chirality (i.e. circular dichroism). Two opposite $\pm\theta$ oblique illumination angles are connected by a simple mirror symmetry in the $(x,y)$ plane and this corresponds to the sequence of transformations detailed as the succession of a flip (a $C_2$ rotation along the $y-$axis) and of a mirror reflection with respect to the $(x,z)$ plane. The result of this sequence is to show, as detailed in the main text, that the optical activity associated with this \textit{extrinsic} 3D chirality induced on the planar chiral system at oblique illumination is reversed for opposite incidence angles $\pm\theta$.}
\label{fig1} 
\end{figure}
which is well recognized in the context of high-resolution chiroptical sensing \cite{FernandezCorbaton2020,Poulikakos2020,ParkPRL2015}, polaritonic physics \cite{Chervy2018,Menon2019}, chiral quantum optics \cite{Zoller2017} and quantum materials \cite{Hubener2020}. In the specific context of light-matter strong coupling studies, such chiral cavities are expected to give rise to chiral polaritonic states that open uncharted paths for driving new, asymmetric, chemical syntheses and material properties \cite{Ebbesen2016,HauglandPRX2020}.

Such cavities however are challenging to design because optical helicity changes sign at each mirror reflection, so that helicity densities are eventually brought to zero through the multiple paths that determine the modal structure of the cavity. Fundamentally, the difficulty stems from the pseudoscalar nature of the optical helicity, itself rooted in the pseudovector nature of the optical spin \cite{Barnett2012}. For this reason, a cavity cannot be chiral without spin-orbit coupling at play, explaining immediately why filling a Fabry-Perot cavity with an optically active material cannot meet the challenge. At the cost of complexity therefore, various schemes have been proposed for realizing helicity-preserving optical cavities, involving for instance intracavity polarization optics \cite{Kastler1970,Rakitzis2014} or elaborate designs ranging from Bragg resonant twisted sculptured thin films \cite{Hodgkinson2000} to optical metamaterial metasurfaces \cite{Plum2015,FernandezCorbatonAPR2020} difficult to scale down to the visible range.

In this Letter, we take another route, simple yet general, by exploiting a deep connection between chirality and optical spin-orbit interactions. We show indeed that optical spin orientations can be locked to intracavity propagation directions when a seed of \textit{planar} (2D) chirality is present inside the cavity. This seed is given by inserting between the two metallic mirrors of a Fabry-Perot cavity a layer of polystyrene made 2D chiral under torsional shear stress. By taking advantage of the extrinsic properties associated with planar chirality under oblique illumination, we demonstrate how the Fabry-Perot cavity can be endowed with an helicity-preserving modal response. This is a clear asset of our chiral cavities that, combined with the simplicity of our approach, immediately makes our systems particularly relevant for applications involving polaritons built on chiral light-matter states hybridized throughout the cavity mode volume.

The dispersive chiroptical features of our cavities are analyzed in the framework of the Jones-Mueller formalism. We start by noting that, unlike three dimensional (3D) chirality associated with optical activity, planar (2D) chirality in its most general expression (without point symmetry like rotational invariance \cite{Drezet2008}) is characterized by polarization transfers from left- to right-handed circular polarization, i.e. from positive to negative helicities, that are flipped when exchanging the enantiomeric form of the 2D chiral structure through which light is transmitted \cite{Fedotov2006,Drezet2008,Plum2009}. 

These peculiar features are most clearly described within the Jones formalism, starting in the circular basis of polarization $|R\rangle,|L\rangle$ with the Jones matrix of a birefringent (linear $LB$ and circular $CB$) and dichroic (linear $LD$ and circular $CD$) optical system: \footnote{We define by $\overline{\bm J}$ Jones matrices expressed in the basis of the circularly polarized states, and $\bm J$ Jones matrices written in the linear polarization basis.}
\begin{align}
 \overline{\bm J}=
  \begin{pmatrix}
 J_{ll} &  J_{lr}\\
  J_{rl} & J_{rr}
 \end{pmatrix} =
   \begin{pmatrix}
\cos\frac{T}{2} + \frac{iC}{T}\sin\frac{T}{2} & -\frac{(iL + L')}{T}\sin\frac{T}{2}  \\
 -\frac{(iL - L')}{T}\sin\frac{T}{2} & \cos\frac{T}{2} - \frac{iC}{T}\sin\frac{T}{2}
 \end{pmatrix}
 \label{CirBJM}
\end{align}
where $T = \sqrt{L^2 + {L'}^2 + C^2}$ for $C=CB - i CD$, $L = LB - iLD$ measured along the linear $|x\rangle,|y\rangle$ polarization axes, and $L' = LB' - iLD'$ along $\pi / 4$-tilted $(|x\rangle\pm|y\rangle) /\sqrt{2}$ linear polarization axes \cite{Arteaga2013,MinghaoPhD}.
The difference $\chi$ between diagonal elements of the Jones matrix is a measure of the optical activity of the system, with a real part proportional to circular dichroism (CD) and an imaginary part associated with circular birefringence (CB) according to:
\begin{align}
\chi = (J_{ll} - J_{rr})/2 = (CD+i CB)\times\frac{\sin \left(T/2\right)}{T}.
\label{3Dc}
\end{align}

For 2D chirality, reciprocity imposes $J_{ll}=J_{rr}$, that is $\chi=0$, from the interconversion of the planar enantiomeric forms of the 2D chiral system when the propagation of the probing light beam is reversed \cite{Hecht1994,Drezet2014,Pham2017}. But in the absence of any point symmetry, the square norm difference $\rho=|J_{rl}|^2 - |J_{lr}|^2$ of off-diagonal elements is non-zero and characterizes 2D chirality through what is known as the circular conversion dichroism (CCD) \cite{Drezet2008,Schwanecke2008,Cao2014}:
\begin{eqnarray}
\rho = (LB\times LD' - LB'\times LD)\times\left(\frac{2\sin \left(T/2\right)}{T}\right)^2
\label{2Dc}
\end{eqnarray}
that stems from the misalignment between $LB$ and $LD$. 

One key point for this work is that CCD associated with 2D chirality couples optical spin with the propagation direction of the light beam, in the sense that reversing the direction of propagation and the helicity as it happens after reflection on one cavity's end-mirror makes the light beam transmitted through the enantiomorphic Jones matrix for which $J_{rl}$ and $J_{lr}$ are exchanged. In striking contrast with 3D chirality, this implies that after one round-trip inside the cavity, this spin-orbit coupling leads to a different left- vs. right-handed circular polarization balance that depends on the initial choice of helicity, as illustrated in Fig. \ref{fig1} (a) and (b).

There is second key aspect associated with a planar chiral system that, viewed at an oblique angle of incidence, yields optical signatures that engage both 2D and 3D chirality. These chiroptical features can be understood by a point group symmetry analysis. A planar chiral system is indeed of $C_{1h}$ symmetry with only one plane of symmetry perpendicular to the optical axis, and thus contrasts with a 3D chiral system having $C_2$ symmetry with an axis of rotation perpendicular to the optical axis. But when observed under oblique incidence $+\theta$, a planar chiral object, described with the Jones matrix $\bm J_{+\theta}$, looses $C_{1h}$ symmetry with no additional $C_2$ symmetry, as clearly seen on the left hand side of Fig. \ref{fig1} (c). As a consequence of the tilt therefore, a flip of the system described by the operation $\bm \Pi_x \bm J_{+ \theta}^T \bm \Pi_x^{-1}$ performed on the Jones matrix written in the linear polarization basis, does neither transform it into its initial configuration by any rotation along the optical $z-$axis (3D chirality) nor into its $(x,y)-$plane mirror symmetrical 2D enantiomer (2D chirality). This means that the system under oblique incidence must be described by a combination of both chiralities with different signatures viewed from both $\pm \theta$ incidence angles. 

The connection between opposite incidence angles $\pm \theta$ can be made most straightforwardly through a $(x,y)-$plane mirror symmetry noted $\sigma_h$ on Fig. \ref{fig1} (c). This simple operation however cannot be directly expressed within the Jones formalism. To do so, we decompose $\sigma_h$ into two successive transformations: one $C_2$ rotation along the $x-$ axis (flipping operation) followed by a $(x,z)-$plane mirror reflection, yielding $\bm J_{+\theta} = \bm J^T_{-\theta} $ where $T$ is the matrix transpose.
Within the circular basis of polarization, this relation becomes $\overline{\bm J}_{+\theta} = \sigma_1 \overline{\bm J}^T_{-\theta}\sigma_1^{-1} $ with $\sigma_1$ the first Pauli matrix. 

The most important consequence of this analysis is that the optical signatures associated with 3D chirality will be reversed under opposite oblique incidence angle while those associated with 2D chirality will be preserved 
\begin{align}
\chi_{\theta} = -\chi_{-\theta} && \rho_{\theta} = \rho_{-\theta} ,
\label{Chidiff}
\end{align}
following the definitions of Eqs. (\ref{3Dc}) and (\ref{2Dc}). As seen, the manifestation of 3D chirality is angle-dependent and as such, is totally different from \textit{intrinsic} 3D chirality generally rotationnaly invariant. This illustrates how \textit{extrinsic} are these 3D chiral features that emerge from 2D chirality at oblique incidence \cite{Plum2009,Markovich2012}. Below, we exploit these relations (\ref{Chidiff}) as a way to characterize the 2D chirality of a system, particularly relevant when the source of planar chirality remains weak.

Our approach to induce 2D chirality inside a Fabry-Perot cavity is to use atactic polymers such as polystyrene \cite{Wulff1989}. When a torsional shear stress is applied to such an atactic polymer, chiroptical features arise in the polymer matrix that are induced by a macroscopic chiral conformation of the chains. We generated the stress inside the polymer matrix by spin-coating clockwise or anticlockwise a thin layer (ca. $150$ nm) of dissolved polystyrene solution (molecular weight of 195 K, diluted $4\%$ in weight in toluene) on a $30$ to $60$ nm thick silver mirror -details are provided in Appendix A. Friction forces the polymer chains to take a macroscopic chiral arrangement close to the surface of the mirror via in-plane spinning of the chains, adopting a macroscopic $C_{1h}$ symmetry of 2D chirality. Far from the surface, the conformation of the polymer chain is not hindered by friction and the chains are simply randomly distributed within the volume \footnote{Despite the fact that the conformation of each chain is chiral, intermolecular chain compensation of chirality prevents any chiroptical signal to be observed macroscopically, a property known as cryptochirality \cite{Mislow1976,Green1984}.} 

Within it, these structural changes correspond to the specific chiroptical features that we analyzed above. Experimentally, the CD signal is measured as the $(0,3)$ coefficient of the cumulated differential Mueller matrix, as explained in Appendix C. The Mueller matrix (MM) itself is acquired on a home-build optical setup that yields calibrated, angle-resolved (Fourier space) MM described in details in Appendix B. Remarkably, the MM gives the possibility to separate linear birefringences and dichroisms from circular ones and thus to measure true planar chiroptical features and artifact free CD \cite{Arteaga2010,MinghaoPhD}.

\begin{figure}[htb]
  \centering{
     \includegraphics{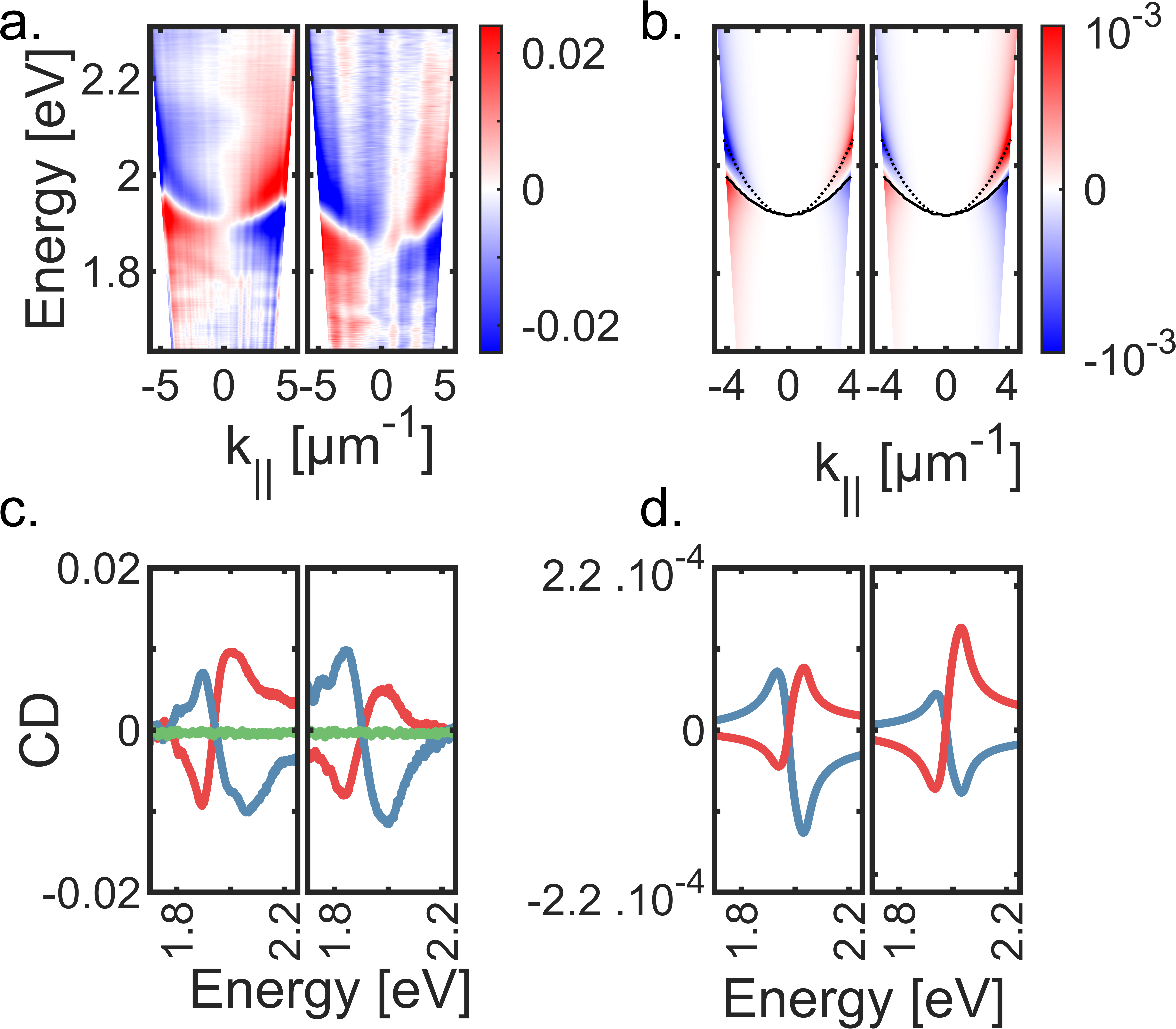}}
    \caption{(a) Measured CD dispersions for both enantiomorphic cavities: clockwise shear stress -left panel- and anticlockwise shear stress -right panel. The $TE$ (continuous line) and the $TM$ (dashed line) modes are superimposed to the measured CD. (b) Simulated CD dispersions for both enantiomorphic cavities for the same clockwise and anticlockwise shear stresses. (c) Associated experimental traces averaged over the two $\pm k_{\parallel}$ angular sub-spaces for both forms where the red and blue traces correspond to averaging performed over the positive and negative sub-spaces, respectively. Once converted in mdeg, the CD values reported correspond to ca. $600$ mdeg. (d) With the same color coding, simulated traces averaged from the simulations shown in (b) over both $\pm k_{\parallel}$ angular sub-spaces.}
\label{fig2}
\end{figure}

The first feature observed for a polymer layer spin-coated on a glass substrate is the absence of CD, a trait expected from the cryptochiral nature of the polymer layer \cite{Mislow1976,Green1984}, and consistent with 2D chirality that does not yield any CD as shown by the green curve in Fig. \ref{fig2} (c). We then form a Fabry-Perot cavity by sandwiching the polymer layer between two Ag mirrors of the same thickness (as explained in Appendix A) and measure its CD in transmission. This time, as seen in Fig. \ref{fig2} (a), clear signatures of CD are observed under oblique illumination. These signatures are remarkable in that they correspond to optically active transverse electric ($TE$) and transverse magnetic ($TM$) modes of the Fabry-Perot cavity. At normal incidence, the opposite helicity between $TE$ and $TM$ modes is a direct consequence of 2D chirality, as explained in Appendix D. At fixed illumination angles where the degeneracy between $TE$ and $TM$ modes is lifted, this yields the bi-signated signatures observed experimentally through the cavity at fixed illumination angles and displayed in Fig. \ref{fig2} (c). Remarkably, as seen in particular in Fig. \ref{fig2} (c) and (d), the contrast  of the $\pm k_{\parallel}$ angularly averaged bi-signated profiles are reversed between opposite enantiomorphic cavities.

\begin{figure}[htb]
  \centering{
    \includegraphics{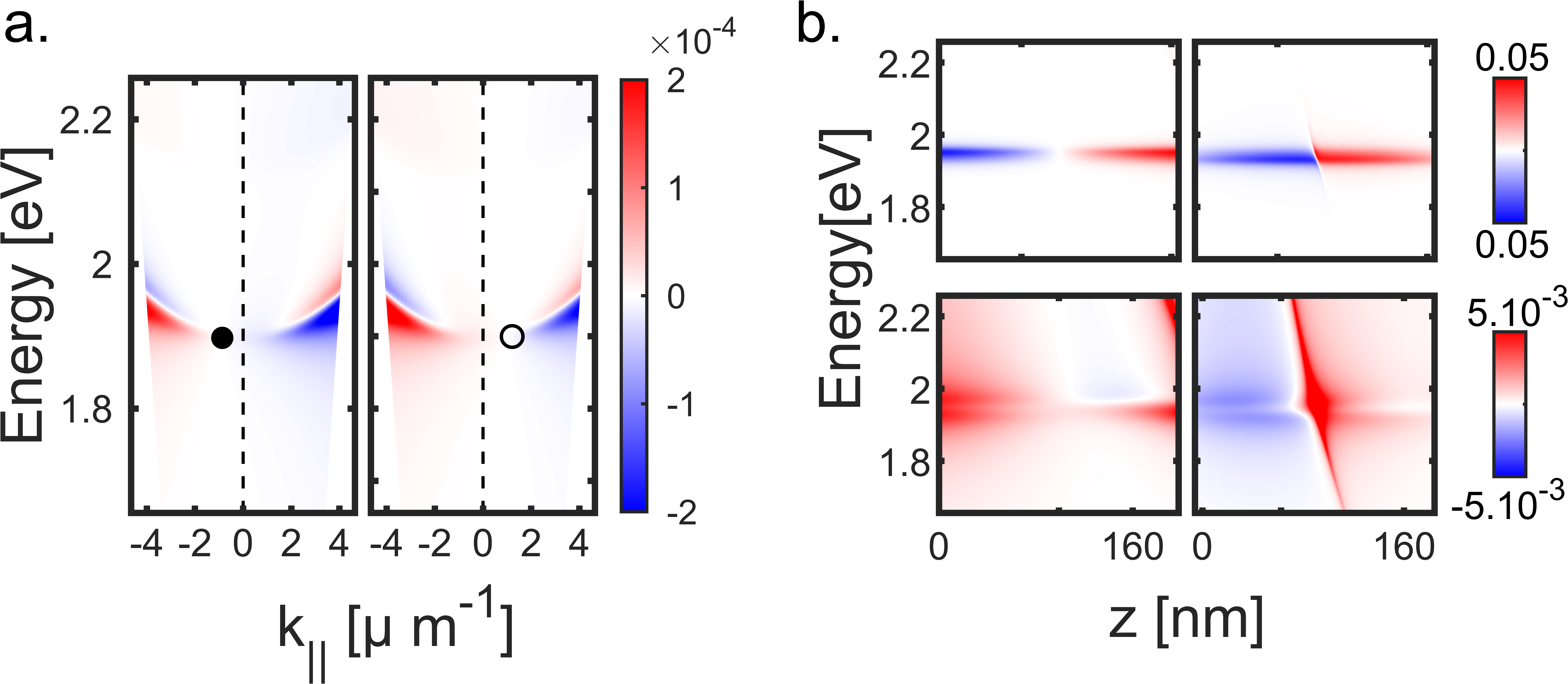}}
    \caption{(a) Global helicity  $\alpha_i(\lambda)$ normalized to the maximum intensity of the intracavity electric field for the $i=TE$ cavity mode calculated for enantiomorphic cavities with clockwise -left panel- and anticlockwise -right panel- shear stresses. For both forms, the in-plane wavevector $k_{\parallel}$ values for zero helicity of the mode are marked with a filled and empty circle, respectively. (b) Intracavity $\delta G_i (z)$ calculated at a chosen $k_{\parallel}=+4~\mu$m$^{-1}$ by placing inside the cavity a 2D chiral medium described by Eq. (\ref{eqn:keff}) (top row) or (bottom row) with the cavity uniformly filled with 3D chiral medium described by a corresponding, constant, $\kappa_{3D}$ (see main text) for both $i=TE$ (left side) and $i=TM$ (right side) modes. Note that the contrast of the bottom row is adjusted for clarity but with values one order of magnitude smaller than those of the top row.}
\label{fig3}
\end{figure}

It is also clear from the data that reveal CD signs exchanged from both sides of the normal incidence that the spin-coated polymer thin film yields a zero CD at normal incidence inside the cavity. As discussed further below, we interpret the tilt of  the whole chiral landscape as an effect of intertwined 2D chirality and extrinsic 3D chirality. This results in the more intense  CD signals observed in Fig. \ref{fig3} (c) in one angular sector in relation with the enantiomorphism of the cavity. 

The properties of our cavities can be simulated using the transfer matrix approach presented in the Appendix E. In this approach, a first approximation describes our polymer film under shear stress as a Pasteur medium, i.e. as a chiral isotropic medium \cite{Lakhtakia1989,Lindell1994}, with the constitutive relations
\begin{eqnarray}
{\bf D}( \textbf{r})&=&\epsilon {\bf E}( \textbf{r})+ i \frac{\kappa}{c}{\bf H}( \textbf{r})  \\
{\bf B}( \textbf{r})&=&-i\frac{\kappa}{c}{\bf E}( \textbf{r})+\mu {\bf H}( \textbf{r})
\end{eqnarray}
where the permittivity ($\epsilon=\epsilon_0\epsilon_r$), the permeability ($\mu=\mu_0\mu_r$) are the usual isotropic parameters ($c^2=1/\epsilon_0\mu_0$) of the polymer medium and $\kappa$ the (complex) parameter associated with its chiral response. This model captures well the experimental features observed in Fig. \ref{fig2} (a) and (c) when describing the chiral response of our material with a $(\theta,\lambda)-$dispersive chiral parameter
\begin{equation}
\kappa_{\rm eff}(\theta,\lambda)=\kappa (\lambda) [ a\times \cos (\theta)+ b\times\sin (\theta) ].
\label{eqn:keff}
\end{equation}

In this effective model, the wavelength dependent complex parameter $\kappa (\lambda)$ is taken to be only weakly dispersive in the visible range, in agreement with the cryptochirality of the polymer itself --see Appendix E. Then the tilt of the 2D chiral material is described by involving the two signatures given in Eq. (\ref{Chidiff}). The parity-even response in $\theta$ associated with 2D chirality is gauged by the $a$ parameter while the 3D chiral parity-odd response in $\theta$ is gauged by the $b$ parameter, driven by the extrinsic 3D chirality emerging from 2D chirality under oblique illumination. By choosing a $b/a\simeq 10$ ratio, the model reproduces well the $(\theta,\lambda)$ dispersion of the MM measured experimentally in strict relation with the enantiomorphism of the cavity, as shown in Fig. \ref{fig2} (b). There is a very good agreement between theory and experiment in the angular evolution of the chiroptical properties of the $TE$ and $TM$ modes, with the bi-signation and the asymmetry in the CD signal measured between the two positive and negative $\pm\theta$ angular sectors observed in Fig. \ref{fig2} (d). These features, both measured and simulated, illustrate the role of planar chirality when present inside a Fabry-Perot cavity.

As we now show, the unique chiroptical properties that planar chirality yields under oblique illumination lead to the possibility to \textit{store} a preferred helicity within a given mode in one cavity round-trip. To demonstrate this, we quantify the chirality of a cavity mode using the metric (used for instance in \cite{FernandezCorbaton2020,Poulikakos2016,Poulikakos2019}) $\delta G ( \textbf{r})=(|\textbf{G}_+ (\textbf{r})|^2 - |\textbf{G}_- (\textbf{r})|^2 )/\sqrt{2}$, where $\textbf{G}_\pm (\textbf{r})= \textbf{E}(\textbf{r})\pm i\eta \textbf{H}(\textbf{r})$ are the Riemann-Silberstein vectors and $\eta$ is the usual impedance of the field. As explained in Appendix E, this impedance within a chiral medium can be simplified to the local difference between left and right electric field intensities \cite{Serdyukov2001}.
We chose this metric because it is directly linked to the optical chiral density and thus directly measures the predominance of one spin-polarized field over the other \cite{Kruining2016}. Integrating $\delta G ( \textbf{r})$ along the $z-$propagation direction inside the chiral film gives the global helicity of the cavity mode $i=TE,TM$ within a $h=z_2 - z_1$ thick layer
\begin{equation}
    \alpha_i(\lambda)=\frac{1}{h}\int^{z_2}_{z_1} \delta G_i ( \textbf{r})dz .
\end{equation}
Those quantity are displayed in panels (a) and (b) in Fig. \ref{fig3} where $\alpha_i(\lambda)$ has been normalized by the field maximum intensity inside the cavity. They demonstrate that the cavity modes defined in our designer Fabry-Perot cavity are characterized by finite helicity densities, whose handedness is opposite in each $\pm\theta$ angular sector. Here too, the combination of 2D and 3D chiralities contributes to the tilt of the chiral landscape and the change of helicity that we expect for the extrinsic 3D chirality is shifted to non-zero incidence angles. The fact that $\alpha_i(\lambda)$ is non-zero along the $i=TE, TM$ modes at normal incidence is a central result of the Letter, with the sign of the helicity of the resonator at normal incidence that depends on the clock/anticlockwise spin-coating direction. This gives our cavities a real potential for exploring resonant strong coupling signatures in chiral polaritonic chemistry and material science. 

The angular evolution of $\alpha_i(\lambda)$ is related to the profile of $\delta G_i$ inside the cavity as shown in Fig. \ref{fig3} (b). The $\delta G_i$ profiles, shown in Fig. \ref{fig3} (b), reveal that  when the cavity is modelled with a 2D chiral layer of the polymer film, chosen here to correspond to a 20\% volume fraction of the cavity, the local helicity of the cavity modes can be enhanced by ca. one order of magnitude in comparison with an intrinsically 3D chiral cavity. \footnote{ The $\kappa_{3D}$ parameter modeling the intrinsically 3D chiral medium  is fixed in such a way that for a chosen angle $\theta_{0}$, it is equal to $\kappa_{eff}(\theta_{0},\lambda)$ with the same parameter value used for modeling the 2D chiral response. We stress that our  modelization based on a  Pasteur medium approach is perfectly appropriate for computing CD dispersions. However our effective treatment of planar chirality within this approach necessarily underestimates the actual strength of $\delta G_i$ inside the cavity.}

In conclusion, we demonstrated that a polymer film on which a chiral stress is imposed can seed planar chirality within a Fabry-Perot cavity. This seed enables a spin-orbit coupling mechanism that shapes, for each round-trip inside the cavity, transverse electric and transverse magnetic modes with a preferred helicity density. Analyzed using the Jones-Mueller formalism, the proposed mechanism for shaping such chiral modes results from the combination between 2D and 3D chiralities under oblique illumination and as such, is a universal mechanism that can be involved in a great variety of system, in particular soft, polymeric media, and over large optical bandwidths. This universality, combined with the simplicity in the implementation, paves the way to exploit such chiral modes in the context of chiral cavity QED \cite{ParkPRL2015,Hubener2020} and polaritonic chemistry \cite{Ebbesen2016}. There, the chiral nature of the polaritonic states that can be created within our cavities yields the core ingredient needed for inducing a new type of selectivity for asymmetric syntheses performed in the regime of strong coupling. This will yield original strategies that we are currently exploring in the endeavor to draw a new landscape for asymmetric chemistry driven by chiral polaritonic states.

\section*{Acknowledgments}

This work was supported by the French National Research Agency (ANR) through the Programme d'Investissement d'Avenir under contract ANR-17-EURE-0024, the ANR Equipex Union (ANR-10-EQPX-52-01), the Labex NIE (ANR-11-LABX-0058 NIE) and Labex CSC (ANR-10-LABX-0026 CSC) projects, the University of Strasbourg Institute for Advanced Study (USIAS) (ANR-10-IDEX-0002-02) and the European Research Council (ERC project no 788482 MOLUSC).


%

\section{Appendix}

\section{A --Sample fabrication}  \label{SI:sample}
The fabrication process for our cavities is simple. We first thoroughly clean a $2.5\times 2.5$ cm$^2$ glass substrate by sonication in a $0.5\%$wt Hellmanex solution in ultra-pure water for 8 minutes. We then sonicate the substrate in a bath of ultra-pure water only, for 30 minutes followed by a sonicated bath in pure ethanol for 30 other minutes. We finally rinse the substrate using a series of 20 dips in a solution of ultra-pure water.

In order to form the cavity itself, we first sputter a layer of silver using an Emitech K575X tabletop sputterer at 60 mA (for 45 s for 30 nm and 95 s for 60 nm Ag layers). On top of the Ag mirror thus sputtered, we spin-coat a solution of dissolved Polystyrene,$4\%$wt in Toluene at 1400 RPM for 2 min to obtain a thin film of 150 nm. The RPM speed was calibrated using profilometry. We finally ``close'' the cavity by sputtering another Ag layer on top, using the same sputter parameters for the same thickness.

\section{B -- Experimental setup}
\label{SI:Mueller}

\begin{figure}[htb]
    \centering
    \includegraphics[width=\linewidth]{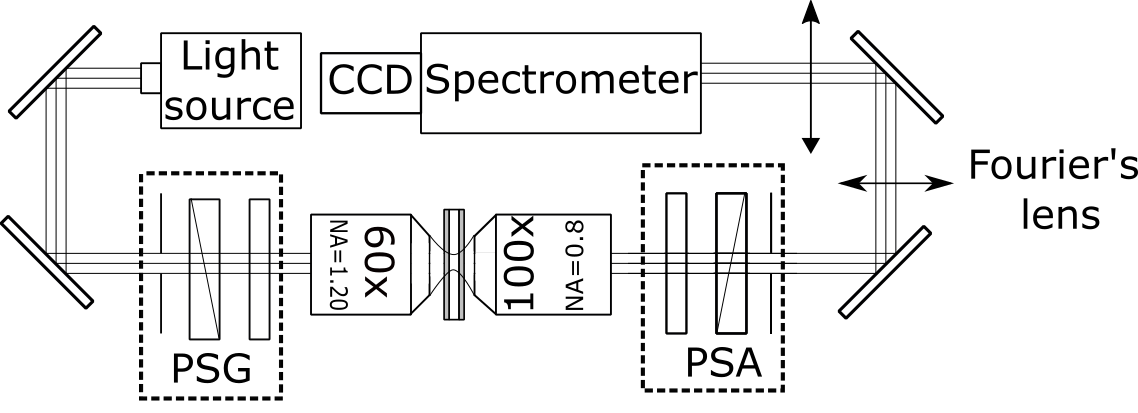}
    \caption{Experimental setup used for the Mueller matrix determination, composed of a polarization state generator (PSG) and analyzer (PSA). The  two lenses give us the ability to image the BFP of the objective, i.e. the Fourier space of our sample.}
    \label{experimental setup}
\end{figure}

Our experimental setup used for the full determination of the Mueller matrix element is schematized in \ref{experimental setup}.
The first part is made of a polarization state generator (PSG) composed of a Glan-Taylor (GT) linear polarizer and a motorized quarter-wave plate. The light beam is injected through the sample using a Nikon ELWD $40\times$ (NA=0.6) objective and collected using a Nikon ELWD $100\times$ (NA=0.9) objective. 
The collected light passes through a polarization state analyzer (PSA) composed of a quarter-wave plate and a linear polarizer (LP). The two linear polarizers were chosen different because of the negative bias induced on the Mueller matrix elements when using a GT within the PSA. 
The last part is made of a lens set at the focal distance from the back focal plane (BFP) of the second objective, associated with a second lens at the entry of a spectrometer (Teledyne Princeton Instrument, SpectraPro HRS-300) for imaging this BFP --referred below as ``Fourier space imaging''. Removing this lens gives us the ability to image the focal plane of the objective on the CCD --referred as ``real space imaging''. Spatially resolved, the spectra are recorded using a PIXIS 1024 CCD camera.

In order to resolve the Mueller matrix for a given wavelength, we build a system of equation linking the measured intensity for a given state of polarization (SOP) to its Mueller matrix (MM) element. The SOP are generated using a carefully chosen combination of angles for the quarter waveplates in both the PSA and PSG. 

Because the Mueller matrix is a $4\times 4$ matrix, there are at least 16 linearly independent equations to solve for our system. Experimentally, we overestimate this minimal set by doing 64 measurements, solving the system by the least-square method. This approach was already detailed by us in \cite{ThomasJPCC2018} and is simply summarized here. 

Let $M_{\rm S}$, $M_{\rm PSA}$, $M_{\rm PSG}$ be the Mueller matrix of the sample, the PSA, and respectively the PSG. We can form the following system of equations represented in the following matrix form:

 \begin{align}
     \bf{S}_{\rm out}&=\bm{M}_{\rm PSA}\bm{M}_{\rm S}\bm{M}_{\rm PSG}\bf{S}_{\rm in}\\
     \bf{S}_{\rm out}&=\bm{M}_{\rm PSA}\bm{M}_{\rm S}\bf{G}
 \end{align}
 The intensity recorded, for one experiment, by the CCD, corresponds to the first element of $S_{out}$, $I_{out}$, which can be expressed as:
 \begin{equation}
    I_{\rm out}=\sum^{4}_{j=1}\sum^{4}_{i=1}m^{\rm PSA}_{1,i}\times g_{i} \times m^{\rm S}_{j-1,i}
    \label{MullerSystem1}
 \end{equation}
where $m^{\rm PSA}_{1,i}$ is the known first line of the PSA Mueller matrix element, $g_{i}$ the known element of the vector resulting from $M_{\rm PSG}.\bf{S}_{in}$, and $m^{\rm S}_{j,i}$ the unknown Mueller matrix element of the sample. We write our set of 64 equations linking the intensity and the Mueller matrix element in the following and most convenient matrix formulation:
\begin{equation}
    {\bf{b}} _{ \rm 64\times 1} = \bm{A} _{ \rm 64 \times 16 } . \bf{X} _{ \rm 16\times 1}
    \label{MullerSystem2}
\end{equation}
where $\bf{b}$ is a containing each intensity of the 64 measurement, and $\bf{X}$ is a vector containing all the Mueller matrix elements. Because we overestimate our system, there is no unique solution but there is a unique solution that minimize the residue $\nu ^2$:
\begin{equation}
    \nu ^2=( {\bf{b}}-\bm{A}{\bf{X}})^T.( {\bf{b}}-\bm{A}{\bf{X}})
    \label{residue}
\end{equation}
The vector that minimize the residue can be expressed as:
\begin{equation}
    {\bf{X}}=(\bm{A}^T\bm{A})^{-1}\bm{A}^{T}\bf{b} . \label{residuesol}
\end{equation}

From Eq. (\ref{residuesol}), one can uniquely define the MM associated with the change of the incident light SOP through the medium. Alone however, the Muller matrix is hardly useful in a chiroptical context. In order to access genuine chiroptical observables, some data filtering is necessary in order to remove any optical artifacts that would prevent us from recovering the relevant chiroptical features of the sample, in particular its circular dichroism.

\section{C -- Data filtering}
\label{SI:data}

From an experimental Mueller matrix, one can develop an algebra, which allows to isolate the CD signal of the  sample and remove artifact signals in the system. This well-known algebra is detailed for instance in \cite{Gil2014}. For our experiments, two different data filtering steps were used when imaging real space vs. Fourier space. 

In the real space, we first identify the polarization responses of the objectives by measuring the MM of $(i)$ a setup with two $\bm{M}_{\rm 40\times}$  and no sample, and $(ii)$ of the setup described in \ref{experimental setup}. We find indeed that experimentally, $\bm{M}_{ \rm 40\times}\simeq \bm{M}_{\rm 100\times}$. One can then remove the response of the objective by noting that when measuring an empty setup, i.e. with no sample, one effectively measures: 
\begin{align}
    \bm{M}_{ \rm empty}&= \bm{M}_{\rm 40\times} \bm{M}_{\rm 40\times}\\
\end{align}
We can then compute $\bm{M}_{\rm 40\times}=(\bm{M}_{\rm empty})^{\frac{1}{2}}$ and remove the responses of our objectives in the real space. In Fourier space the same procedure is applied.

The second filtering step consists in removing the contribution of the glass substrate of the sample from our experimental Mueller matrix. To do so, we first measure the response in both real and Fourier spaces of the cleaned substrate, $\bm{M}_{glass}$. Then, we can determine the Mueller matrix of the sample alone without the glass substrate contributions by using the following serial Mueller decomposition:
\begin{equation}
    \bm{M}_{\rm S}=\bm{M}_{\rm glass}\bm{M}_{\rm FP}
\end{equation}
where $\bm{M}_{\rm FP}$ is the Mueller matrix of the Fabry-Perot cavity (without substrate) which can be rewritten as $\bm{M}_{\rm FP}=\bm{M}_{\rm glass}^{-1} \bm{M}_{\rm S}$.

The third  filtering step is to decompose the previously obtained Mueller matrix using the Cloude decomposition \cite{Cloude1990}. The goal of the method is to give an estimation of the equivalent non-depolarizing Mueller matrix, known as Mueller-Jones matrix, necessary for the last filtering step below. Following \cite{Gil2014,Savenkov2009}, Cloude decomposition consists in computing the $4\times 4$ hermitian coherency matrix $\bm{T}$, which will have the following matrix elements:
\begin{align*}
    &t_{11}=\frac{1}{4}(m_{00}+ m_{11}+m_{22}+m_{33})\\
    &t_{12}=\frac{1}{4}(m_{01}+ m_{10}-i(m_{23}-m_{32}))\\
    &t_{13}=\frac{1}{4}(m_{02}+ m_{20}-i(m_{31}-m_{13}))\\
    &t_{14}=\frac{1}{4}(m_{03}+ m_{30}-i(m_{12}-m_{21}))\\
    &t_{22}=\frac{1}{4}(m_{00}+ m_{11}-m_{22}-m_{33})\\
    &t_{23}=\frac{1}{4}(m_{12}+ m_{21}-i(m_{30}-m_{03}))\\
    &t_{24}=\frac{1}{4}(m_{13}+ m_{13}-i(m_{02}-m_{20})\\
    &t_{33}=\frac{1}{4}(m_{00}- m_{11}+m_{22}-m_{33})\\
    &t_{34}=\frac{1}{4}(m_{23}+ m_{32}-i(m_{10}-m_{01}))\\
    &t_{44}=\frac{1}{4}(m_{00}- m_{11}-m_{22}+m_{33})\\
\end{align*}
The coherency matrix can be computed from any given experimental matrix. By considering that any depolarizing Mueller matrix M can be considered as a convex sum of non-depolarizing Mueller-Jones matrix, denoted $\bm{M}_{Ji}$, one can link the eigenvalues $\lambda_i$ of the coherency matrix to  $\bm{M}_S$  by the following:
\begin{align}
     \bm{M}=\sum^{3}_{i=0}=\lambda_i \bm{M}_{Ji}\\
    \bm{ M}_{Ji}=\bm{A}.(\bm{J_i} \otimes \bm{J_i}^* )\bm{A}^{-1}
\end{align}
 where $\bm{J}_i$ is the  Mueller-Jones matrix associated to $\bm{M}_{Ji}$ and $\bm{A}$ is the passage matrix that can be written as 
 \begin{equation}
     \bm{A}=\begin{pmatrix}
     1 & 0& 0& 1\\
     1 & 0& 0& -1\\
     0 & 1 &1& 0\\
     0 & -i &i& 0\\
     \end{pmatrix}
 \end{equation}
 
We then can rank the $\bm{M}_{Ji}$ in terms of their respective weight. Generally, this decomposition is dominated by the first term, ie $\lambda_0>>\lambda_1,\lambda_2,\lambda_3 $, and one can consider $\lambda_0 \bm{M}_{J0}$ as a good estimate of the non-depolarizing Muller matrix associated to $\bm{M}_S$. From this estimate, one can directly extract the CD following \cite{Arteaga2013}. We first compute the cumulated differential Mueller matrix $\bm{L}_m$ as
\begin{align}
\bm{L}_m = &{\rm ln}( \bm{M}_{\rm J0}) \\
=& \frac{1}{2} ( \bm{L} - \bm{G} \bm{L}^T \bm{G} )
\end{align}
where G is the Minkowski tensor G=diag(1,-1,-1,-1) and ${\rm ln}$ the matrix logarithm. In this manner, the CD simply corresponds to the $\bm{L}_m(0,3)$ matrix element.

\section{D --Bi-signated CD signals for TE and TM modes}  \label{SI:bisig}

In order to explain the bi-signated CD signal for the $TE$ and $TM$ modes measured through the cavity under oblique illumination, we look at the helicity of the transmitted beam under $TE$ and $TM$ polarizations at normal incidence, where the two associated modes are degenerated. Within the Stokes-Mueller formalism, $TE$ and $TM$ modes are expressed by Stokes vectors as $\mathbf S_{TM} = (1,-1,0,0)^T$ and $\mathbf S_{TE} = (1,1,0,0)^T$. Since our system at normal incidence is a 2D chiral system, it is simply described by a Jones-Mueller matrix given by
\begin{equation}
\bm M_{2D} = 
\begin{pmatrix}
 1 & -a_1& -a_2& 0 \\ -a_1 & d_1 & 0 & b_2 \\ -a_2& 0 & d_2 & -b_1\\ 0 & -b_2 & b_1 & d_3
\end{pmatrix},
\end{equation}
and therefore yielding the TE and TM transmitted Stokes vectors:
\begin{equation}
\mathbf S_{TM}^{\rm out} =  \bm M_{2D} \cdot \mathbf S_{TM} = 
\begin{pmatrix}
 1+a_1 \\ -a_1-\alpha \\ -a_2 \\  b_2
\end{pmatrix}
\end{equation}
\begin{equation}
\mathbf S_{TE}^{\rm out} =  \bm M_{2D} \cdot \mathbf S_{TE} = 
\begin{pmatrix}
 1-a_1 \\ -a_1+\alpha \\ -a_2 \\  -b_2 
\end{pmatrix}.
\end{equation}

The $S_3$ elements $\pm b_2$ for each transmitted Stokes vector are opposite. This implies that the helicities associated with the $TE$ and $TM$ modes, degenerated at normal incidence, are opposite. When the optical activity emerges at an oblique angle accompanied by a lifting of degeneracy, the CD for the $TE$ and $TM$ branches will have opposite sign accordingly.

\section{E -- Chiral transfer matrix}
\label{SI:Tmat} 

In the liquid crystal community, the usual approach followed for simulating transmission spectra is the Berreman matrix formalism \cite{Berreman1972}. But this method suffers from the rise of singularities in certain specific cases \cite{Wu2018}.
To overcome this problem, we model, in a first approximation, our polymer film under shear stress as a typical Pasteur medium, i.e. as an isotropic chiral medium. We however introduce a spatially dispersive response of the chirality parameter of the medium in order to mimic its real response associated with extrinsic/intrinsic chirality.   

In a Pasteur medium, one derives the constitutive equations as \cite{Lindell1994}: 
\begin{equation}
\begin{split}
{\bf{D}}=\epsilon {\bf{E}}+ i\frac{\kappa}{c}{\bf{H}}
\end{split}
\hspace{1cm}
\begin{split}
 {\bf{B}}=-i\frac{\kappa}{c}{\bf{E}}+\mu {\bf{H}}
\end{split} ,
\end{equation}
where the permittivity ($\epsilon$),  the permeability ($\mu$) and  the chiral parameter ($\kappa$) are usual isotropic parameters. The source-free wave equation for such a Pasteur medium is given by \cite{Lindell1994}:
\begin{equation}
\bm{\nabla}^2{\bf{E}}-i\omega\frac{\kappa}{c}{\bf{\nabla}} \wedge {\bf{E}}+ i\omega (\frac{\kappa}{c} \eta \mu - \epsilon \mu) {\bf{E}}=0 .
\end{equation}
In this case, the eigenstates are circularly polarized plane waves and the electromagnetic field inside the chiral medium can be written as a superposition of right and left polarized waves going in the forward and backward directions:
\begin{align}
\begin{split}
{\bf{E}} = & {\bf{E}}_{+R}  e^{-i( {\bf{k_+}} . {\bf{r}} + \omega t)}+{\bf{E}}_{+L} e^{ i({\bf{k_+}}.{\bf{r}} -\omega t)}+ 
\\ 
& {\bf{E}}_{-R} e^{-i( {\bf{k_-}} .{\bf{r}}+ \omega t)}+{\bf{E}}_{-L} e^{i( {\bf{k_-}} . {\bf{r}}-\omega t)}
\end{split} ,
\end{align}
where $R$ and $L$ denote the  right-going and left-going plane waves. These eigenstates feel the chirality of the medium as a standard medium without electromagnetic coupling, i.e.:
\begin{align}
    {\bf{D}}_\pm=\epsilon_\pm {\bf{E}}_\pm &\quad& {\bf{B}}_\pm =\mu_\pm{\bf{H}}_\pm ,
\end{align}
where one can derive $\epsilon_\pm$ and $\mu_\pm$ from the constitutive parameters and the wavenumber associated with each polarization state \cite{Lindell1994}:
\begin{align}
\mu_\pm=\mu\pm\frac{\kappa}{c}\sqrt{\frac{\mu}{\epsilon}} && \epsilon_\pm = \epsilon \pm \frac{\kappa}{c}\sqrt{\frac{\epsilon}{\mu}}&
 &\quad k_\pm=\omega(\sqrt{\epsilon\mu}\pm\frac{\kappa}{c}) .
\end{align}

Our approach is similar to \cite{Jaggard1992} and to other transfer matrix computations.
Once the field is characterized in the chiral medium, the field continuity equation can be written in a convenient matrix form ${\bf{E}}_{n-1} =A_{n-1,n} {\bf{E}_{n}}$ where $A$ is a bloc symmetric $4\times 4$ matrix linking the right and left polarized electric field going forward and backward from the $(n)$ layer to the $(n-1)$ layer which are gathered in the following quadrivector ${\bf{E}}_{\rm n}$:
\begin{equation}
\begin{pmatrix}
E_{+R}\\E_{-R}\\E_{+L}\\E_{-L}
\end{pmatrix}_{n-1}
=\begin{pmatrix}
\bm{a}_T & \bm{a}_R \\ \bm{a}_R & \bm{a}_T
\end{pmatrix}
\begin{pmatrix}
E_{+R}\\E_{-R}\\E_{+L}\\E_{-L}
\end{pmatrix}_{n},
\end{equation}    
where:
\begin{align}
	\bm{a}_T=&
	\begin{pmatrix}
		\frac{\eta_r+1}{4}(1+\frac{cos(\theta_{+,n})}{cos(\theta_{+,n-1})}) & \frac{\eta_r-1}{4}( 1- \frac{cos(\theta_{-,n})}{cos(\theta_{+,n-1} )}) \\
		\frac{\eta_r-1}{4}( 1- \frac{cos(\theta_{+,n})}{cos(\theta_{-,n-1} )}) & \frac{\eta_r+1}{4}(1+\frac{cos(\theta_{-,n-1})}{cos(\theta_{-,n-1})})
	\end{pmatrix}
	&\quad &\\
	\bm{a}_R=&
	\begin{pmatrix}
		\frac{\eta_r+1}{4}(1-\frac{cos(\theta_{+,n})}{cos(\theta_{+,n-1})}) & \frac{\eta_r-1}{4}( 1+ \frac{cos(\theta_{-,n})}{cos(\theta_{+,n-1} )})\\
		\frac{\eta_r-1}{4}( 1+ \frac{cos(\theta_{+,n})}{cos(\theta_{-,n-1} )}) & \frac{\eta_r+1}{4}(1-\frac{cos(\theta_{-,n})}{cos(\theta_{-,n-1})})
	\end{pmatrix}
\end{align}
with $\eta_r=\frac{\eta_{n-1}}{\eta_{n}}$ the ratio between the usual wave impedance in their respective layer and $\theta_{\pm,n}$ the of the wave associated with the left or right polarized field that one can recover by imposing continuity of the phase at the interface, according to:
\begin{equation}
\theta_{\pm,n}=asin(\frac{k_{\pm,n-1} sin(\theta_\pm)}{k_{\pm,n}}) .
\label{angle}
\end{equation}

To take into account the phase gained by the electric field inside one layer, we introduce the $P_n$ matrix:
\begin{equation}
\bm{P_n}=
\begin{pmatrix}
e^{-ib_{+}} & 0 & 0 & 0\\
0 & e^{-ib_{-}} & 0 & 0\\
0 & 0 & e^{ib_{+}} & 0\\
0 & 0 & 0 & e^{ib_{-}}\\
\end{pmatrix}
\label{Eq_phase}
\end{equation}
where $b_{\pm}=k_{\pm,n}d_ncos(\theta_{\pm,n})$.
With this, the final total transfer matrix, $T_{tot}$, can be written as the following product of matrix:
\begin{equation}
	\bm{T}_{tot}=\bm{A}_{0,1} \bm{P}_{1} \bm{A}_{1,2}...\bm{P}_N \bm{A}_{N-1,N} .
\end{equation}

\begin{figure}[htb]
    \centering
    \includegraphics{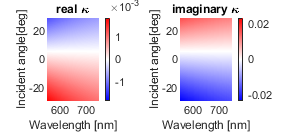}
    \caption{ Real -panel (a)- and imaginary -panel (b)- parts of $\kappa_{eff}$.}
    \label{fig1SI}
\end{figure}

Finally, in order to measure the total field intensity transmitted by our sample, we set $E_{+L,N}=E_{-L,N}=0$ and compute:
\begin{equation}
 	\begin{pmatrix}
	 	E_{+R} \\ E_{-R}
	 \end{pmatrix}_{out}
 	=
 	\begin{pmatrix}
 		t^{tot}_{11} & t^{tot}_{12}\\
 		 t^{tot}_{21} & t^{tot}_{22}
	 \end{pmatrix}^{-1}
 	\begin{pmatrix}
	 	E_{+R} \\ E_{-R}
	 \end{pmatrix}_{in} .
\label{transmittion_equation}
\end{equation}
From this equation, one can easily calculate the total transmission of our sample. Moreover, probing the sample's medium with four linearly independent Stokes vectors leads to compute the Mueller matrix of the sample using the chiral parameter indicated in Fig.\ref{fig1SI}. In order to model this, we assume a resonance in the UV that is optically active. This resonance yields a non-zero imaginary part for the chiral parameter, which fixes the dispersive nature of our polymer, even far from the resonance through the (broad-band) Kramers-Kronig relation. 

\begin{figure}[htb]
    \centering
    \includegraphics{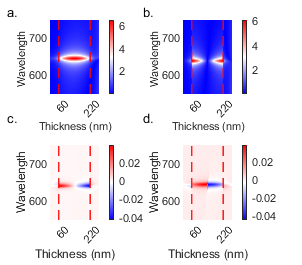}
    \caption{ $TE$ (a) and $TM$ (b) intracavity mode electric field intensities evaluated at an angle of -35 deg ($k_{\parallel}=-5~\mu$m$^{-1}$). Corresponding $\delta G_{TE}(\textbf{r})$ (c) and $\delta G_{TM}(\textbf{r})$ (d). The thin film boundaries are indicated with red dashed lines.}
    \label{fig2SI}
\end{figure}

Using the previously calculated matrix $A_{n,n+1}$  and $P_n$ and their relative $z-$positions inside the layer, one can compute the electric field intensity at any point within our multilayer system. With this, we compute the the Riemann-Silberstein vectors inside the chiral medium as defined in the main text:
\begin{equation}
    \textbf{G}_\pm (\textbf{r})= \textbf{E}(\textbf{r})\pm i\eta \textbf{H}(\textbf{r})
\end{equation}
where $\eta$ is the usual impedance of the medium. By assuming that $H=\frac{i}{\eta}E$, we write:
\begin{equation}
	\delta G (\textbf{r})=|\textbf{G}_{+}(\textbf{r})|^{2}-|\textbf{G}_{-}(\textbf{r})|^{2}.
\end{equation}
Our transfer matrix simulation allows monitoring $\delta G(\textbf{r})$ with respect to its position along the $z-$axis for both TE and TM modes. The raw results are presented in Fig. \ref{fig2SI}.

\begin{figure}[htb]
    \centering
    \includegraphics{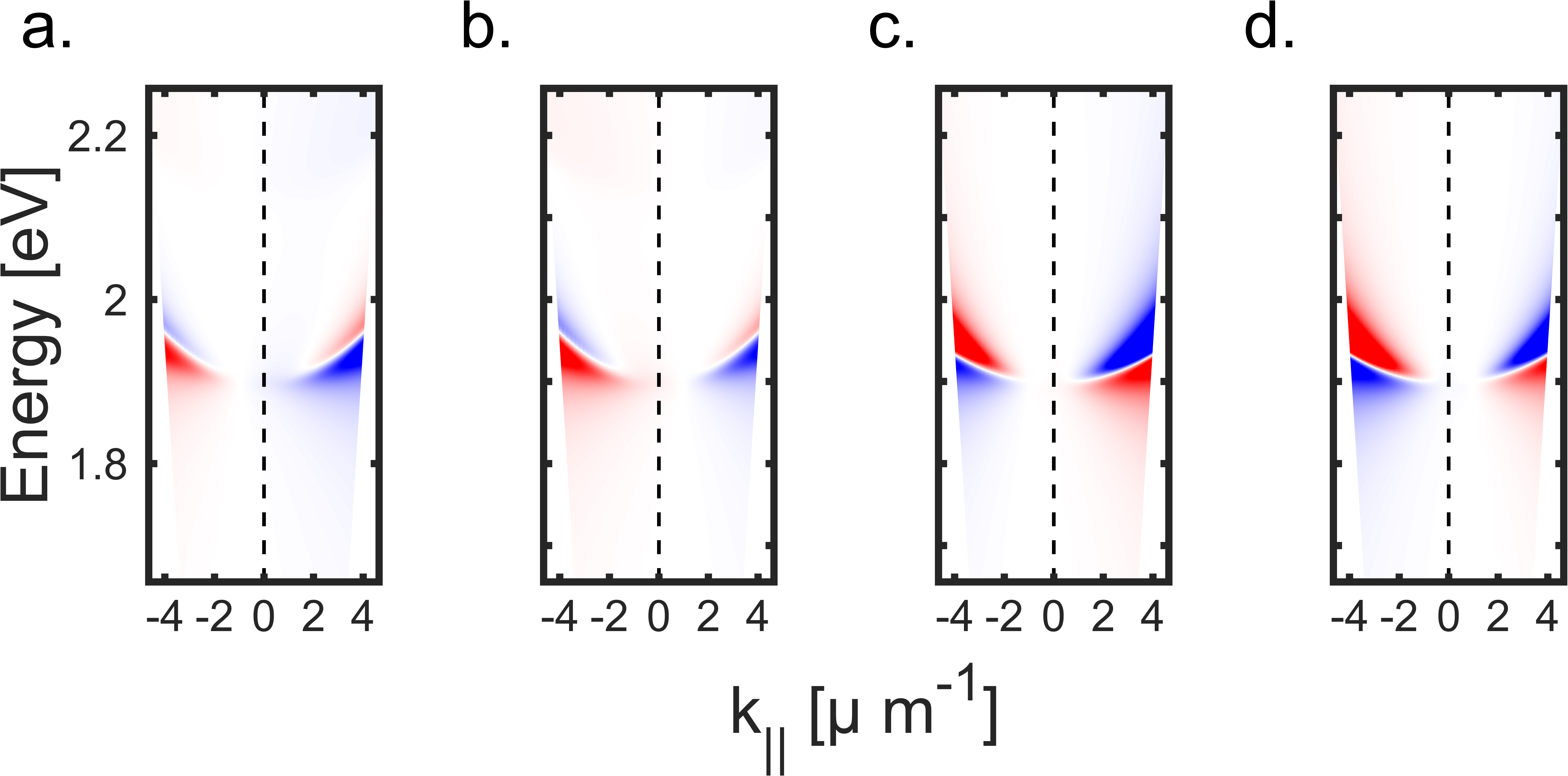}
    \caption{(a) Global helicity $\alpha_i(\lambda)$ for the $i=TE$ (a)-(b) and $i=TM$ (c)-(d) cavity mode calculated for enantiomorphic cavities with clockwise - (a)and (c)- and anticlockwise -(b)and (d)- shear stresses. }
 \label{SI_fig4}
\end{figure}

This variable leads us to monitor the local helicity of the field inside the cavity. In addition, we compute the predominant helicity of the mode by integrating along the $\hat{z}$ axis, denoting it by $\alpha_\lambda=1/h\int^{z_2}_{z_1} \delta G ( \textbf{r})dz$. The two simulations presented on Fig. 3 (a) in the main paper   are obtained by changing the enantiomeric form associated with the extrinsic 3D chirality coming from the  planar chiral structure. The value for the TE and TM modes are indicated in \ref{SI_fig4}\\
 
A key result is the change of the preferential helicity of the field at normal incidence as we change the enantiomeric form of the planar structure. The point where the helicity flips sign  is determined by the relative strength between 2D chirality  (parameter $a$ in the model) and 3D chirality (parameter $b$ in the model). In our simulations, we choosed $b/a=10$, as discussed in the main text.

\bibliography{biblio}

\end{document}